# Classifying Proposals of Decentralized Autonomous Organizations using Large Language Models

Christian Ziegler[1,2][0000-0002-6509-4349] and Marcos Miranda[2] and Guangye Cao[3] and Gustav Arentoft[2] and Doo Wan Nam[2,4]

[1] Technical University of Munich, Arcisstraße 21, 80333 Munich, Germany
[2] StableLab, Sin Min Lane #06-76, Midview City, Singapore
[3] University of Michigan, Ann Arbor, Michigan, USA
[4] Johns Hopkins University, Maryland, Baltimore, USA

**Abstract.** Our *study demonstrates the effective use of Large Language Models (LLMs) for automating the classification of complex datasets. We specifically target proposals of Decentralized Autonomous Organizations (DAOs), as the classification of this data requires the understanding of context and, therefore, depends on human expertise, leading to high costs associated with the task. The study applies an iterative approach to specify categories and further refine them and the prompt in each iteration, which led to an accuracy rate of 95% in classifying a set of 100 proposals. With this, we demonstrate the potential of LLMs to automate data labeling tasks that depend on textual context effectively.*

## 1      Introduction

Decentralized Autonomous Organizations (DAOs) are information systems with different functions that either mediate interactions between humans and blockchains or operate as a completely autonomous system with features that enable storage, transaction of value, voting mechanisms, autonomous execution of governance decisions in a decentralized environment (Hassan & Filippi, 2021; Rikken et al., 2023; Schillig, 2022)

Governance in DAOs is implemented with proposals that have different phases such as pre-discussions, forum discussions, voting, and implementation. These proposals can change any aspect of the DAO, such as allocating funds, changing risk parameters of a Decentralized Finance (DeFi) application, upgrading the protocol, changing the rule for governance, or engaging in partnerships with other DAOs or companies.

DAOs can take many different forms that considerably change how governance works. For example, off-chain product and service DAOs do not run any protocol updates, investment-focused DAOs do not change risk parameters, and a networking-focused community DAO will decide on many more partnership proposals than an on-chain product and service DAO (Ziegler & Welpe, 2022).

Prominent DeFi DAOs such as Aave, Uniswap, Balancer, Safe, Compound, Lido, and Arbitrum have decided on 1645 proposals and discussed those on 3742 topics on the forums from July 2020 to December 2023, which highlights the frequent use of governance proposals to run the DAOs. In the same timeframe, 231442 proposals were created by 35238 DAOs on Snapshot alone.

This vast amount of very different proposals makes it very difficult for researchers to analyze the impact of proposals on, for example, DAO performance since a standard proposal that makes a minor adjustment of a parameter from a DeFi protocol only has a minor effect on the DAO is in large scale indistinguishable from a high impact proposal that makes a change to the core protocol. Currently, researchers have to manually analyze proposals to then use them in a subsequent analysis.

This manual process for classification is very time-consuming and, therefore, very costly on a large scale. At a scale of more than 231442 proposals in Snapshot alone, this task is also unfeasible even for a more extensive research team.

Therefore, we formulate the following research questions:
- **RQ1**: What categories of DAO proposals exist, and what is their prevalence?
- **RQ2**: How can researchers automatically classify DAO proposals?

Contrary to previous and related research that uses data augmentation to enhance training to be more diverse, this approach performs a fully-fledged classification of context-related datasets.

The remainder of this paper is structured as follows. Section 2 presents related work on data and text augmentation using Language Models (LMs) and LLMs. Then, Section 3 introduces our design science methodology with Peffers et al. (2007) and Nickerson et al. (2013). In Section 4, we present our three resulting artifacts. Finally, in Section 5, we conclude.

## 2 Method

In this work, we use the Design Science Research Method (DSRM) proposed by Peffers et al. (2007) as a guideline and Nickerson et al. (2013) to create a taxonomy of proposals in DAOs. Following Peffers et al. (2007), we first identify the problem, state our motivation, and justify the value of a solution. Second, we define the objectives for a solution by inferring the solution's objective from the stated problems. Third, we follow Nickerson et al. (2013) to build a taxonomy of proposals while continuously improving the categories, LLM parameters and the LLM prompt. During each iteration, we perform a demonstration by classifying proposals and an evaluation by comparing the accuracy of the LLM classification against manual classification by delegates.

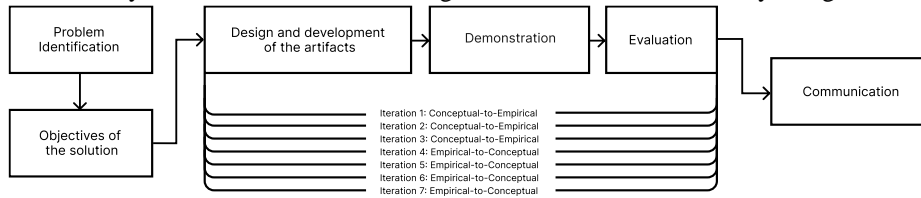

**Fig. 1.** Design Science Research Method adapted from Peffers et al. (2007).

### 2.1 Problem definition

There exists no broadly accepted categorization of DAO proposals. As DAO governance is very diverse in its tasks, a joke proposal to "buy a lamborghini" is indistinguishable from a significant proposal that, for example sets a parameters in the governed DeFi protocol without manual review of the proposal by a human. This manual classification is a very time-consuming and, therefore, expensive task but required for further research into the effectiveness of governance in DAOs.

### 2.2 Objectives of the solution

We, therefore, aim to define categories that cover the whole spectrum of proposal types in DAOs that can be used for further, more insightful research on DAO governance. In addition, we require a reliable and highly accurate classification method that utilizes LLMs to automate classification. Lastly, we want the outcome of this research to be highly re-usable for other researchers.

### 2.3 Design and Development, Demonstration and Evaluation

In total, we created three artifacts: Proposal categories, the LLM prompt, and the LLM framework parameters. We start by creating the proposal categories using the iterative taxonomy-building methodology of Nickerson et al. (2013) with the intent of creating proposal categories and not a fully-fledged taxonomy. According to the iterative approach of Nickerson et al. (2013), we first define our meta-characteristics as Categories to differentiate different types of proposals of DAOs usable for automatic classification with an LLM.

Next, we define our ending conditions by adopting the subjective ending conditions of Nickerson et al. (2013). We, therefore, require our categories to be concise, robust, comprehensive, extendible, and self-explanatory. In addition to the subjective ending conditions, we also define objective ending conditions that deviate from the objective ending conditions of Nickerson et al. (2013), as we are not building a taxonomy but merely categories. Their ending conditions primarily refer to the splitting, merging, addition, or deletion of dimensions or categories in each iteration, asking researchers to do another iteration when one of the named events happens. We define two ending conditions: no new category has been created in the current iteration, and no modifications have been made to existing categories. Lastly, we define an ending condition closely related to the other two artifacts, LLM prompt and LLM framework parameters: At least 90% classification accuracy during the evaluation of the current iteration.

Following the basic setup, we start performing our iterations. In Table 1, we show our seven iterations, complete with the demonstration and evaluation as required by Peffers et al. (2007). We abbreviated conceptual-to-empirical as c-e and empirical-to-conceptual as e-c. The iterations were performed as follows: We first conceptualized new categories or derived categories from DeFi DAOs for the category artifact. The conceptualization was performed with the help of six delegates of DAOs. Delegates in DAOs receive voting rights from tokenholders and vote in their names. They actively participate in governance and get paid for this activity. Therefore, their insights for categorization were invaluable. The observations of the empirical-to-conceptual rounds stem from the DAOs aave.eth, arbitrumfoundation.eth, balancer.eth, comp-vote.eth, lido-snapshot.eth, safe.eth, uniswap. More specifically, from their respective Snapshot spaces and discourse forums. The iterations for the LLM prompt and LLM parameters were performed using a literature review about parameters and trial and error for the LLM prompt until the LLM replied with a valid JSON that included a classification of the input proposals. In total, we performed three conceptual-to-empirical and four empirical-to-conceptual iterations.

**Table 1.** Iterations according to Nickerson et al. (2013).

| Iteration | Design and Development | Demonstration | Evaluation |
| --- | --- | --- | --- |
| 1 c-e | Conceptualization of the Dimensions, initial taxonomy, initial prompt, initial llm parameters | 10 Proposals classified | Classification Accuracy 10% |
| 2 c-e | Changes to prompt and LLM parameters to get consistent results; Makes Categories much more verbose, c-e through balancer, uniswap, safe, lido, aave, compound, arbitrum | 10 Proposals classified | Classification Accuracy 20% |
| 3 c-e | Changes to LLM prompt for consistent results in JSON format; Classification of 20 Proposals by 6 different delegates under the lead of a researcher; Workshop to improve categories with 6 delegates; Update of Categories | 20 Proposals classified | Classification Accuracy 20% |
| 4 e-c | Interview with 5 delegates and 4 DAO operators; Add very specific categories such as gauges, whitelisting wrapped tokens, gas rebates, managing airdrops to their respective categories; Manually classify 100 proposals with delegates and researcher; Update of Categories | 100 Proposals classified | Classification Accuracy 62% |
| 5 e-c | Add very specific categories derived from the 100 proposals that were misclassified; Update Categories | 100 Proposals classified | Classification Accuracy 84% |
| 6 e-c | Manually Classify 100 Discourse discussions by a researcher and the delegates; Add specific categories derived from the 100 proposals that were misclassified; Update Categories | 100 Proposals classified; 100 discourse discussions classified | Classification Accuracy 92% proposals, 77% discussions |
| 7 e-c | Add very specific categories derived from the 100 discourse discussions, and the 100 proposals; No updates to categories | 100 Proposals classified; 100 discourse discussions classified | Classification Accuracy 95% proposals, 95% discussions |

## 2.4 Communication

We shared our research findings through various channels to reach a broad audience. For the scientific community, we compiled this detailed research paper. Additionally, we crafted a blog post featured in the news section of StableLab, a delegate company, making our research accessible to a broader audience. To further enhance the visibility of both the research paper and the blog post, we actively promoted them across several social media platforms.

# 3 Resulting Artifact

In this chapter, we present the three resulting artifacts, starting with the categorization of DAO proposals, then showing and explaining the LLM prompt, and lastly, we present the LLM parameters and explain them. The focus of this chapter is that other researchers can directly extract the three artifacts from this research paper and classify their existing dataset on DAO proposals using our approach, therefore improving the specificity of their research.

## 3.1 Categorization of DAO Proposals

**Treasury and Asset Management (TAM) -** Oversee the DAO's own treasury and assets. This encompasses decisions concerning the security, investment, diversification, and financial reporting of the DAO's own assets, as well as managing associated risks. In this context, the DAO is the asset owner, and these assets form part of its treasury. This also includes potential airdrops that the DAO could receive.

**Protocol Risk Management (PRM) -** Manage operational, technical, liquidity, and other risks related to the protocol or the assets held within the protocol. It also includes Risk and Parameter Reports and Updates related to managing the protocol risk. Responsibilities include adjusting protocol parameters (also referred to as risk parameters), enlisting or delisting assets, ensuring the safety of value and assets locked in the protocol, identifying potential attack vectors, addressing risks inherent to protocol operations, rectifying technical vulnerabilities, and navigating specific ecosystem or contextual threats (which encompasses regulatory and legal risk management).

**Protocol Features and Utility (PFU) -** Enhance and oversee the protocol's functionalities and utility. Responsibilities encompass developing and deploying new code, implementing protocol upgrades, launching new products, deploying new gauges, implementing liquidity mining programs, implementing protocol incentives, expanding the core protocol to additional chains and Layer 2 solutions, and managing the utility of the protocol's native token(s).

**Governance Administration and Framework Management (GAFM) -** Covers proposals that direct the governance process by refining and standardizing the governance framework, rules, processes, templates, and timelines. It also includes Governance

Reports and Updates regarding to Governance. Responsibilities encompass defining roles, managing voting mechanisms and parameters, setting eligibility criteria for voting power, whitelisting tokens into voting escrows and governance contracts, managing Snapshot space and configurations, and determining quorum thresholds. Additionally, this vertical addresses proposals that create or iterate upon processes for onboarding and offboarding roles and entities vital to governance operations, such as service providers, facilitators, working groups, and councils.

**Budget Allocation and Work Management (BAWM)** - Covers proposals that allocate the DAO's budget to internal DAO projects, tasks, and roles requiring execution or oversight. These initiatives may be singular projects or ongoing operations. It includes Community Updates from service providers that keep the DAO informed on various activities, excluding Governance Reports, Financial Reports, and Risk and Parameter Reports. It identifies service providers, individuals, or teams who take on these responsibilities and carry them out according to the defined Scope of Work and designated deliverables. This ensures the efficient utilization of resources in alignment with the DAO's strategic goals and operational demands. This encompasses the allocation and management of duties and work related to marketing, operations, software development, and risk and financial management.

**Partnerships and Ecosystem Development (PED)** - Encompasses proposals aimed at driving external growth via strategic partnerships and multifaceted strategies. The focus is on bolstering the DAO/protocol ecosystem through the formation and maintenance of partnerships, launching educational campaigns, overseeing grant programs, engaging in regulatory and legal activism, contributing resources to external foundations that contribute to wider ecosystem development, and allocating budgets to external software development projects that build upon the core systems of the protocol. Additionally, it emphasizes initiatives designed to keep or/and draw more participants into the protocol ecosystem, such as making airdrops and making users whole in front of eventualities. Also Includes activities that foster community spirit and engagement, such as meetups, social media interactions, content creation, and other forms of outreach that do not explicitly fall under marketing or partnerships. Also covers Informative materials and discussions aimed at improving the knowledge base of the DAO's community members regarding blockchain, the protocol's features, and best practices within the space. Furthermore, includes recognizing and managing the contributions that do not directly impact governance but contribute to the health and growth of the DAO's ecosystem, such as voluntary community moderation, unsolicited user-generated content, and miscellaneous feedback.

**Miscellaneous (MISC)** - Comprehensive umbrella for activities, requests, and contributions that fall outside the predefined governance verticals or are tangential to governance yet are contribute to the DAO's operations. It includes support requests for technical assistance and user troubleshooting, addresses general inquiries about the DAO and its operations, and translation of important documentation to other languages.

For all figures (Appendix), only the most predominant proposal category was counted. When, for example, a proposal has a rating of 0.9 for GAFM and 0.8 for BAWM, it will only show up as GAFM in the charts. From Figure 5 we can see that both lending protocols aave.eth and comp-vote.eth primarily have PRM proposals, while the exchanges uniswap and balancer.eth have mostly PFU proposals and balancer.eth has more proposals in total because of the many gauges.

### 3.2 The Prompt

During the creation of the prompt, we needed to fulfill several requirements. First, categories needed to be explained in the prompt, and their abbreviation must be directly stated as they are required for the output. Second, the full text of the proposal body and the title must be included in the prompt in a way that the body or title of the proposal can not be mistaken as part of the instructions. Third, clear instructions are needed on what data the LLM should compute. Fourth, our goal was to extract as much information as possible from the proposal using a single prompt, as the additional computational power required to compute more data points from a proposal within one prompt is minimal compared to re-running all proposals again. Fifth, we require the LLM to provide a clear reasoning as to which category was chosen for a proposal. We do this to be able to iterate on wrongly classified proposals. Six, we require the output to be in a valid JSON format so that we can directly store the output in a relational database and further use it from there.

In addition to the categorization, we prompted the LLM to check if the personal wealth of the voter is affected, come up with its own categorization, provide the perceived risk for the dao of this proposal, extract the total cost and revenue, perform emotion detection and sentiment analysis, rate the professional structure of the proposal and check if the given proposals is a linked to a previous proposal and if it is a recurring proposal.

### 3.3 LLM Model (Parameters)

For the parameter selection, we are limited to those available in the API reference of OpenAI as we use GPT-4 for this study. Four of the parameters are OpenAI specific, while three are generally available in most LLMs. We first start with the particular OpenAI parameters:

- **model**: specifies the AI model to be used for generating responses. In this case, "gpt-4-0613" indicates a specific version of the GPT-4 model.
- **messages**: This is an array of message objects representing the conversation history. Each message is a dictionary with two keys:
- **role**: Can be either "user" or "assistant," indicating who is sending the message.
- **content**: The actual text of the message. In this case, the prompt would be the variable containing the user's input.
- **max_tokens**: This determines the maximum length of the response. The value 500 indicates that the response can be up to 500 tokens long. A token can be a word or part of a word, so this doesn't directly translate to a specific number of words.

The following three parameters change the outcome of the prompt by introducing randomness, penalizing repetition, or the likelihood that new topics are introduced:

- **temperature**: This controls the randomness of the response. A temperature of 0 means there is no randomness; the model will always give the most likely response based on its training. Higher temperatures lead to more varied and sometimes less predictable responses (OpenAI, 2023; Xue et al., 2023). We set this to 0.
- **frequency_penalty**: This reduces the model's tendency to repeat the same line of text. A penalty of 0 means there's no adjustment for repetition. We set this to 0.
- **presence_penalty**: This influences the model's tendency to introduce new topics or concepts. A penalty of 0 means the model isn't encouraged or discouraged from introducing new topics. We set this to 0.

## 4 Conclusion

Our study aims to improve DAO research to understand their features better and open up more ways of evaluating them. We set the goal of finding categories for DAO proposals, finding their prevalence, and automatically classifying them. Our primary motivations for these goals are the diverse proposals that govern DAOs that can not be universally used for research and the inaccessibility of manually classifying large quantities of data.

To reach this goal, we have performed a design science research method according to Peffers et al. (2007) with seven iterations in the steps of design and development, demonstration, and evaluation. Within these iterations, we performed three conceptual-to-empirical rounds and four empirical-to-conceptual rounds. We draw from the experience of delegates who have voted on hundreds of proposals and evaluate the outcome of each demonstration by comparing the results of the LLM to our manual classification. In our last iteration, we reach an accuracy of 95% over 200 data points.

With this method, we successfully created three artifacts. First, the categories for proposals: Treasury and Asset Management, Protocol Risk Management, Protocol Features and Utility, Governance Administration and Framework Management, Budget Allocation and Work Management, Partnerships and Economic Development, and Miscellaneous. Second, a prompt for LLMs to automatically classify DAO proposals in the given categories. Third, a set of parameters for GPT-4.0 from OpenAI to receive consistent results in the classification.

We contribute to theory in two ways. First, by adding to the understanding of LLMs and their use for automatic data classification. Second, by providing a tested categorization of DAO proposals that can be used in future research. Furthermore, we contribute to practice by providing the complete prompt, parameters, and categories so that any researcher and practitioner can replicate our findings.

Our work acts as a starting point for in-depth research on DAO proposals. We foresee that quantitative and qualitative research on each proposal type will increase, leading to a deeper understanding of the dynamics and decision-making processes within DAOs. Future research can potentially assess the effectiveness of each proposal category and find bottlenecks in DAO governance.
Using the artifacts, we classified 1614 proposals and 3572 discourse discussions from Aave.eth, arbitrumfoundation.eth, balancer.eth, comp-vote.eth, lido-snapshot.eth, safe.eth, Uniswap.

# References

1. Anaby-Tavor, A., Carmeli, B., Goldbraich, E., Kantor, A., Kour, G., Shlomov, S., Tepper, N., & Zwerdling, N. (2020). Do Not Have Enough Data? Deep Learning to the Rescue! *Proceedings of the AAAI Conference on Artificial Intelligence*, *34*(05), 7383–7390. https://doi.org/10.1609/aaai.v34i05.6233
2. Brown, T. B., Mann, B., Ryder, N., Subbiah, M., Kaplan, J., Dhariwal, P., Neelakantan, A., Shyam, P., Sastry, G., Askell, A., Agarwal, S., Herbert-Voss, A., Krueger, G., Henighan, T., Child, R., Ramesh, A., Ziegler, D. M., Wu, J., Winter, C., . . . Amodei, D. (2020). *Language Models are Few-Shot Learners.* https://doi.org/10.48550/arXiv.2005.14165
3. Chen, J., Yang, Z., & Yang, D. (2020). *MixText: Linguistically-Informed Interpolation of Hidden Space for Semi-Supervised Text Classification.* https://doi.org/10.48550/arXiv.2004.12239
4. Chong, D., Hong, J., & Manning, C. D. (2022). *Detecting Label Errors by using Pre-Trained Language Models.* https://doi.org/10.48550/arXiv.2205.12702
5. Fadaee, M., Bisazza, A., & Monz, C. (2017). Data Augmentation for Low-Resource Neural Machine Translation. In R. Barzilay & M.-Y. Kan (Eds.), *Proceedings of the 55th Annual Meeting of the Association for* (pp. 567–573). Association for Computational Linguistics. https://doi.org/10.18653/v1/P17-2090
6. Hassan, S., & Filippi, P. de (2021). Decentralized Autonomous Organization. *Internet Policy Review*, *10*(2). https://doi.org/10.14763/2021.2.1556
7. Kumar, V., Choudhary, A., & Cho, E. (2020). *Data Augmentation using Pre-trained Transformer Models.* https://doi.org/10.48550/arXiv.2003.02245
8. Lee, K., Guu, K., He, L., Dozat, T., & Chung, H. W. (2021). *Neural Data Augmentation via Example Extrapolation.* https://doi.org/10.48550/arXiv.2102.01335
9. Nickerson, R. C., Varshney, U., & Muntermann, J. (2013). A method for taxonomy development and its application in information systems. *European Journal of Information Systems*, *22*(3), 336–359. https://doi.org/10.1057/ejis.2012.26
10. OpenAI. (2023). *OpenAI - API reference.* https://platform.openai.com/docs/api-reference/chat
11. Peffers, K., Tuunanen, T., Rothenberger, M. A., & Chatterjee, S. (2007). A design science research methodology for information systems research. *Journal of Management Information Systems*, 45–77.
12. Rikken, O., Janssen, M., & Kwee, Z. (2023). The ins and outs of decentralized autonomous organizations (DAOs) unraveling the definitions, characteristics, and emerging developments of DAOs. *Blockchain: Research and Applications*, *4*(3), 100143. https://doi.org/10.1016/j.bcra.2023.100143
13. Sahu, G., Rodriguez, P., Laradji, I. H., Atighehchian, P., Vazquez, D., & Bahdanau, D. (2022). *Data Augmentation for Intent Classification with Off-the-shelf Large Language Models.* https://doi.org/10.48550/arXiv.2204.01959
14. Schillig, M. A. (2022). Decentralized Autonomous Organizations (DAOs) under English law. *Law and Financial Markets Review*, *16*(1-2), 68–78. https://doi.org/10.1080/17521440.2023.2174814
15. Wei, J., & Zou, K. (2019). *EDA: Easy Data Augmentation Techniques for Boosting Performance on Text Classification Tasks.* https://doi.org/10.48550/arXiv.1901.11196
16. Wu, X., Lv, S., Zang, L., Han, J., & Hu, S. (2018, December 17). *Conditional BERT Contextual Augmentation.* http://arxiv.org/pdf/1812.06705v1

17. Xue, T., Wang, Z., & Ji, H. (2023). *Parameter-Efficient Tuning Helps Language Model Alignment.* https://doi.org/10.48550/arXiv.2310.00819
18. Yoo, K. M., Park, D., Kang, J., Lee, S.-W., & Park, W. (2021). *GPT3Mix: Leveraging Large-scale Language Models for Text Augmentation.* https://doi.org/10.48550/arXiv.2104.08826
19. Ziegler, C., & Welpe, I. M. (2022). A Taxonomy of Decentralized Autonomous Organizations. In *ICIS 2022 Proceedings*.

# Appendix – Figures

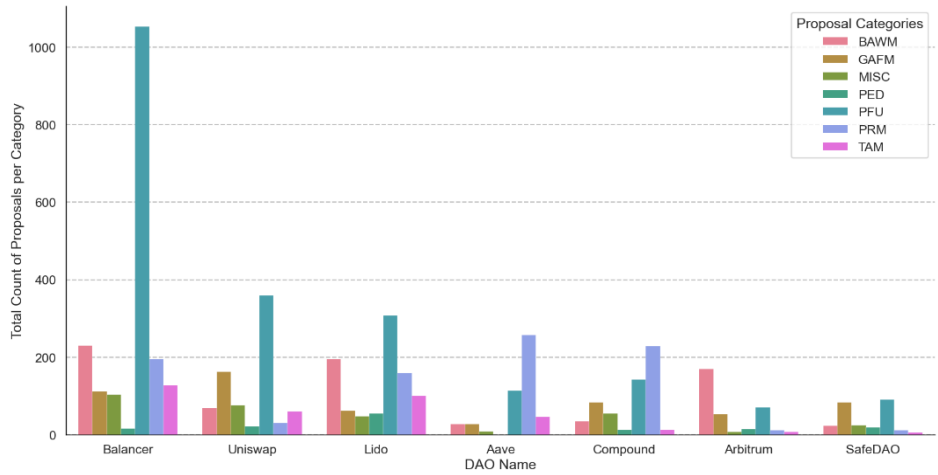

**Fig. 2.** Proposal category occurrence by DAO in total numbers.

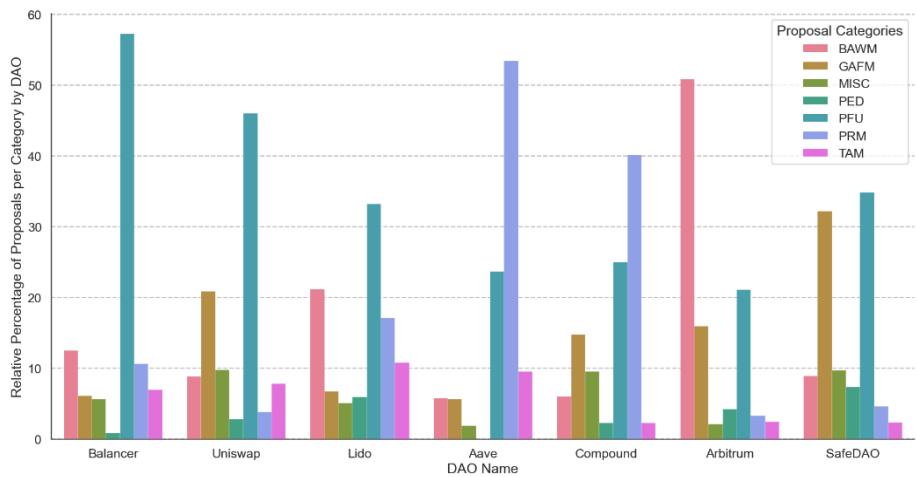

**Fig. 3.** Proposal category occurrence in relative percentages by DAO.

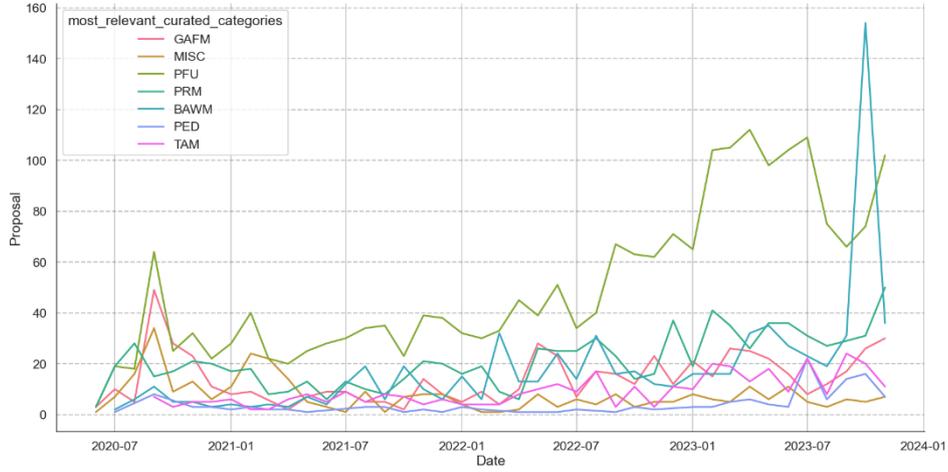

**Fig. 4.** Proposal count in the selected DAOs over time.

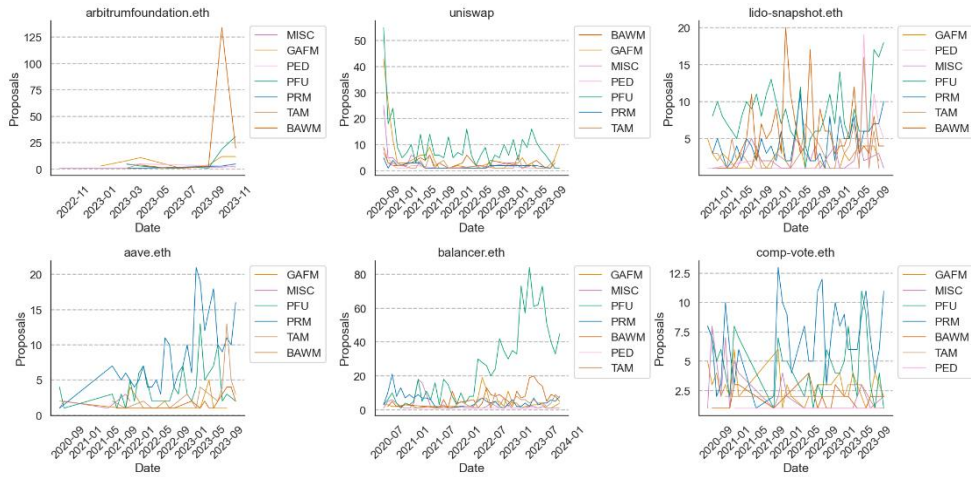

**Fig. 5.** Proposal count by DAO and category over time.

# Appendix - Prompt

The following is the title and description of a Proposal for a Decentralized Autonomous Organization (DAO).
Please analyze the following title and body of the proposal and classify it using the categories and their explanation that are listed afterward
TITLE: {Proposal Title}.
BODY: {Proposal Body}.
BODY END
You can ONLY choose from the following curated categories:
Categories: [TAM, PRM, PFU, GAFM, BAWM, PED, MISC]
Explanation: {Categories Explained}
Also answer the following question:
Does the proposal affect the personal stake or wealth of the voters? (true/false)
Use the following JSON template with example values to answer using a percentile how certain you are with your evaluation.
Replace y with at least one category shortcut, z with a reasoning, x with a number from 0 to 1. Additonally, for llm_categories, come up with at least one top level category that would fit the proposal in order for a researcher to later do clustering on them
Also perform a sentiment analysis and provide the values in the sentiment arrays.
Convert all price ranges to their average. Convert abreviations like K=Thousand, M=Million to the responding full number.
ALWAYS respond with a valid json for python with the following structure:
{
 'personal_wealth_affected: false,'
 'most_relevant_curated_categories: ,'
 'clear_reasoning: z,'
 'categories: {'
  'TAM: x,'
  'PRM: x,'
  'PFU: x,'
  'GAFM: x,'
  'BAWM: x,'
  'PED: x'
  'MISC: x'
 },
 'llm_categories: ,'
 'risk_for_dao: number,'
 'total_cost: number $currency or false,'
 'total_revenue: number $currency or false,'
 'emotion_detection: [{example_emotion: 0.x, etc.}],'
 'fine_grained_sentiment: [{example_sentiment: 0.x, etc.}],'
 'professional_proposal_structure_score: number,'
 'previous_proposal: bool or id,'
 'is_recurring_proposal: bool'
}